\documentclass{iucr}              

     \paperprodcode{a000000}      
     \paperref{xx9999}            
     \papertype{SC}               

     \paperlang{english}          
     \journalcode{J}              
     \journalyr{2010}
     \journaliss{1}
     \journalvol{59}
     \journalfirstpage{000}
     \journallastpage{000}
     \journalreceived{0 XXXXXXX 0000}
     \journalaccepted{0 XXXXXXX 0000}
     \journalonline{0 XXXXXXX 0000}

\begin{document}                  



\title{3D Spatially Resolved Neutron Diffraction from a Disordered Vortex Lattice}

\author[a]{Xi}{Wang}
\author[a]{Helen A.}{Hanson}
\cauthor[a]{Xinsheng Sean}{Ling}{Xinsheng$\_$Ling@brown.edu}

\author[b]{Charles F.}{Majkrzak}
\author[b]{Brian B.}{Maranville}

\aff[a]{Department of Physics, Brown University, Providence RI 02912, \country{USA}}
\aff[b]{NIST Center for Neutron Research, Gaithersburg MD 20899, \country{USA}}



\shortauthor{Wang, Hanson, Ling, Majkrzak \& Maranville}







\maketitle                        



\begin{abstract}
The vortex matter in bulk type-II superconductors serves as a prototype system for studying the random pinning problem in condensed matter physics. Since the vortex lattice is embedded in an atomic lattice, small angle neutron scattering (SANS) is the only technique that allows for direct structural studies. In traditional SANS methods, the scattering intensity is a measure of the structure factor averaged over the entire sample. Recent studies in vortex physics have shown that it is highly desirable to develop a SANS technique which is capable of resolving the spatial inhomogeneities in the bulk vortex state. Here we report a novel slicing neutron diffraction technique using atypical collimation and an areal detector which allows for observing the three dimensional (3D) disorder of the vortex matter inside an as-grown Nb single crystal.
\end{abstract}

\section{Introduction}

A major goal in the field of vortex physics is to determine the nature of the long-range order in the ground state of the Abrikosov vortex line lattice (Abrikosov, 1957).  It has been theoretically predicted that in systems with weak random pinning, a new phase of matter known as a Bragg glass exists in the vortex state (Chudnovsky, 1989; Nattermann, 1990; Giamarchi \& Le Doussal, 1995).  In contrast to a perfect crystal with a delta-function structure factor, the Bragg glass theory predicts a power-law shaped structure factor (Chudnovsky, 1989; Giamarchi \& Le Doussal, 1995;  Bogner \emph{et al.}, 2001), similar to a 2D crystal at finite temperatures with no random pinning (Imry \& Gunther, 1971; Nelson \& Halperin, 1979). Since the vortex lattice is embedded in an atomic lattice, neutron scattering (Furrer \emph{et al.}, 2009) is the only technique for direct structural studies. There have been several attempts (Klein \emph{et al.}, 2001; Daniilidis \emph{et al.}, 2007;  Laver \emph{et al.}, 2008) to observe the Bragg glass phase, but no conclusive evidence has been found even though there are compelling arguments (Klein \emph{et al.}, 2001) for why one would not be able to measure it directly.

In particular, the vortex states in bulk Nb samples have been studied extensively (Schelten \emph{et al.}, 1971; Christen \emph{et al.}, 1975; Lynn \emph{et al.}, 1994; Rosov \emph{et al.}, 1994; Ling \emph{et al.}, 2001; Forgan \emph{et al.}, 2002; Daniilidis \emph{et al.}, 2007; Laver \emph{et al.}, 2008).  These SANS measurements revealed the bulk properties of the interior of the samples. However, it has been recently observed that the vortex matter in a Nb single crystal can be spatially inhomogeneous (Hanson \emph{et al.}, 2010). Since the SANS measurement is based on an averaged signal, it would be difficult to discern the nature of disorder in the vortex matter. It is thus necessary to develop a tool to spatially resolve the vortex matter structure from different locations in a bulk superconductor.

By using a thin ribbon-shaped beam, as prepared on the Advanced Neutron Diffractometer/Reflectometer (AND/R) at the NIST Center for Neutron Research (NCNR), we find that it is possible to spatially resolve distinct, locally coherent regions of the vortex lattice within the sample (similar to crystallographic grains, with a small angular mismatch between regions), transverse to the direction of the magnetic field (Hanson \emph{et al.}, 2010). Here, we report an improvement on this method which allows us to probe variations of the the in-plane vortex matter structure factor along the direction of the applied magnetic field. We demonstrate the effectiveness of this method in revealing the splitting of a Bragg peak along the magnetic field direction of the vortex lattice in a Nb single crystal.

\section{Sample Properties}

The sample is an as-grown Nb single crystal (cylinder shaped) with its $\left\langle 111 \right\rangle$ direction along the cylindrical axis, used in the previous study (Hanson \emph{et al.}, 2010). It has a diameter of \mbox{$12.1$ mm}, a height of \mbox{$10.1$ mm}, and the weight is \mbox{$9.69$ g}. The residual resistivity ratio (RRR) is $32$. By using the ac magnetic susceptometry technique (Ling \& Budnick, 1991), we have mapped out the phase diagram and observed a weak peak effect in this sample. The characteristic field values at \mbox{$T=4.2$ K} are the lower critical field \mbox{$H_{c1}=0.100$ T}, the upper critical field \mbox{$H_{c2}=0.490$ T}, and the field at the peak of the peak effect \mbox{$H_{p}=0.400$ T}. 

We apply the magnetic field parallel to the $\left\langle 111 \right\rangle$ direction of the Nb crystal. The vortex lattice is prepared with a zero-field cooling procedure followed by thermal annealing and a small ac magnetic field perturbation. The magnetic field (\mbox{$H=0.140$ T}) is applied after the sample is cooled through the transition temperature of \mbox{$T_c=9.2$ K} down to \mbox{$T=4.0$ K}. We thermally anneal the vortex matter by slowly increasing the temperature (in increments of \mbox{$\Delta T=0.5$ K}) to \mbox{$T=7.5$} K, just below the transition temperature, \mbox{$T_{c2}$($H=0.140$ T)}, and then slowly cooling back down to \mbox{$T=4.0$ K} (here $\Delta T=1.0$ K). Finally, we perturb the vortex matter by applying an ac magnetic field (sawtooth-function) with an amplitude of \mbox{$0.001$ T} and a frequency of \mbox{$1.0$ Hz} for $5.0$ minutes provided by a Helmholtz coil outside the superconducting magnet.  We find that this procedure is optimal for forming a stable vortex state with the highest order in this sample.

\section{Conventional Neutron Scattering on Vortex Matter}

\begin{figure}
\includegraphics{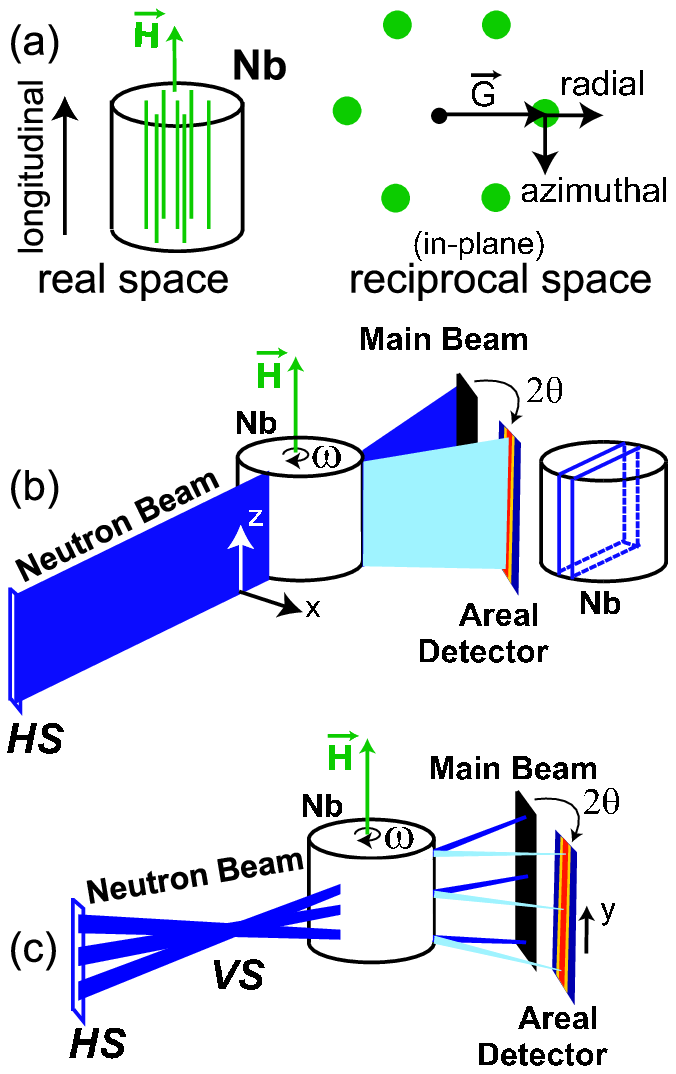}
  \caption{(a) A sketch of our bulk Nb single crystal in a magnetic field with the corresponding structure of an ideal vortex lattice in reciprocal space. (b-c) The scattering geometry for the neutron diffractometer. The magnetic field is applied parallel to the cylindrical axis (the $\left\langle 111\right\rangle$ direction of the Nb crystal). (b) The sample and magnetic field are rotated together by an angle $\omega$ with respect to the incident neutron beam. By using horizontal slits, $HS$, the incident neutron beam is collimated transverse to the flux lines.  By shifting the sample along the $x$ direction, different slices of the sample are probed. An areal detector is placed at the angle $2\theta$ which is determined by the applied magnetic field. The slice in the Nb crystal shows the part of the sample exposed to neutrons. (c) By adding vertical collimation, $VS$, after the horizontal slit in the neutron path in conjunction with an areal detector, we can physically distinguish the scattered neutrons from different vertical sections of the sample. }
	\label{SlicingGeometry}
\end{figure}

In a neutron scattering experiment, the Bragg condition (Bragg \& Bragg, 1913) occurs when the scattering vector $\vec{Q}$ equals the reciprocal lattice vector $\vec{G}$, where $|\vec{G}| = \frac{2\pi}{d} = \frac{4\pi}{\lambda}\sin\theta$. Here $\lambda$ is the neutron wavelength, $\theta$ is the Bragg angle, and $d$ is the distance between lattice planes. In vortex physics (Tinkham, 1996), the applied magnetic field, $H$, determines the value of $d$ through the relationship
\begin{equation}
d=\frac{\sqrt{3}}{2}a=\sqrt{\frac{\sqrt{3}\Phi_{\circ}}{2H}}
\label{dequation}
\end{equation}
where $a$ is the vortex lattice constant and \mbox{$\Phi_{\circ}=2.07\times 10^{-15}$ Wb} is the flux quantum.  The sketch of an ideal vortex lattice in our sample is shown Fig.\ \ref{SlicingGeometry}(a).

The AND/R scattering geometry is shown in Fig.\ \ref{SlicingGeometry}. The parameters of the setup are described elsewhere (Dura \emph{et al.}, 2006) but some important aspects are reiterated here. There are two slits in front of the sample, each one defined by two borated aluminum plates. With both slit widths set to \mbox{$1.0$ mm}, the incident neutron beam has an angular spread of \mbox{HWHM$ = 0.042^\circ$}. A wide third slit is placed directly behind the sample to reduce the background scattering that reaches the areal detector. A fourth slit can be placed directly in front of the detector to further reduce background noise and improve the instrument resolution. In Fig.\ \ref{SlicingGeometry} (b), only the second slit is plotted.

The azimuthal direction (perpendicular to $\vec{G}$) is probed by rotating the sample and the magnetic field (and the associated reciprocal lattice) while keeping the detector (and the associated scattering vector) fixed in space at an angle $2\theta$. The radial direction (along $\vec{G}$) is measured by rotating the sample (and the magnetic field) and neutron detector in concert, always moving the detector by $2\Delta \omega$ for every sample movement of $\Delta \omega$.  This procedure keeps the scattering vector at a fixed azimuthal angle but varies the magnitude with respect to the reciprocal lattice vector of the sample. By scanning the scattering vector through the Bragg condition in both azimuthal and radial directions we can map out the in-plane shape of the Bragg peak.

In previous SANS experiments on vortex matter, the scattering intensities along the azimuthal and radial directions are usually resolution-limited. The trade-off for the high resolution of a reflectometer is a relatively low neutron flux compared to typical SANS instruments. The flux restricts our study to the structure of the vortex matter in the low field region of the phase diagram where the scattering is strong. 

Additionally, in the typical SANS geometry, one illuminates as much of the sample as possible to increase the scattering intensity. The measurement is an averaged signal over the bulk sample. One particularly relevant aspect of the AND/R geometry is the width of the beam compared to the sample diameter (here \mbox{$\le 1.0$ mm} vs. \mbox{$12.1$ mm}). Since only a small slice of the sample is exposed to the neutron beam, we can shift the position of the sample relative to the neutron beam along the $x$ direction ( labeled in Fig.\ \ref{SlicingGeometry}(b)) and study the vortex matter at different sample locations (Hanson \emph{et al.} 2010). The position of the neutron beam relative to the sample ($x$ value) is located by scanning through the neutron absorbers (Cd masks) around the sample holder and the uncertainty of the $x$ value is \mbox{$\pm1.0$ mm}. The data in Fig.\ \ref{data}(a) is a three-dimensional plot of the integrated intensity versus azimuthal angle $\omega$ versus $x$ value, where the origin of $x$ represents the center of the sample. The intensities have been scaled for the different scattering volumes. From Fig.\ \ref{data}(a), it is clear that the Bragg peaks are spatially inhomogeneous (physical implications are discussed in Hanson \emph{et al.}, 2010). 

\section{Three Dimensional Structure Mapping}

The ability to scan the sample along the $x$ direction (see Fig. \ref{SlicingGeometry}(b)) allows us to map the Bragg peak from the sample edge to the sample center and study the spatial dependence of the vortex matter structure. Since the detector is sensitive to the impact of neutrons in two dimensions, by adding vertical collimation to the instrument, we can observe the orientational disorder along the length of the flux lines. 

\begin{figure}
\includegraphics{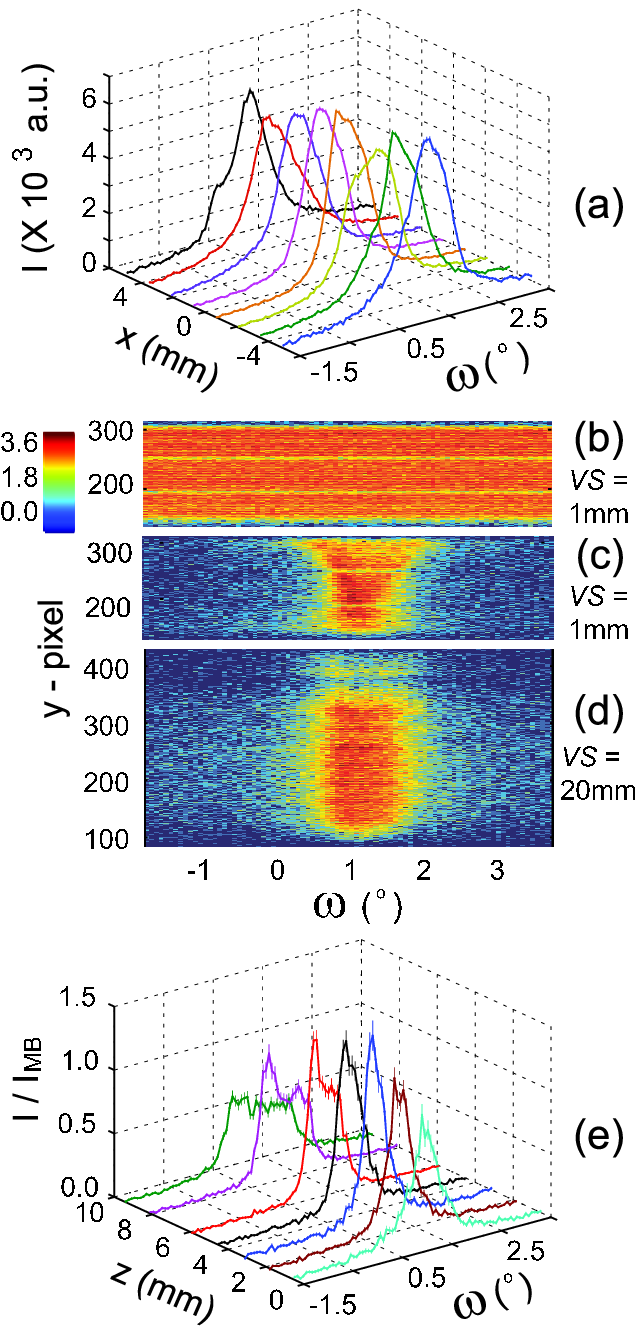}
\caption{(a) The integrated intensity of the scattered neutrons from different horizontal slicing sections of the vortex matter as a function of $\omega$. The integrated intensity is calculated by summing over the areal detector and it has been scaled for the different scattering volumes. The $x$ value represents the distance to the center of the sample. (b) A two dimensional color plot (on a logarithmic scale) of the main beam ($2\theta=0^{\circ}$) with a vertical slit of \mbox{$1$mm}, revealing the dependence of the main beam on the monochromator. (c) A two dimensional color plot of the Bragg peak ($2\theta=0.23^{\circ}$) at \mbox{$x=-0.75$ mm} with a vertical slit of \mbox{$1$ mm},  revealing the raw vertical scattering intensity from the vortex matter. (d) A two dimensional color plot of the Bragg peak ($2\theta=0.23^{\circ}$) at \mbox{$x = -0.75$ mm} with a wide open vertical slit (\mbox{$20$ mm}). (e) The integrated intensity from separating (c) into seven sections. Due to the vertical variations in the main beam, the integrated intensity is normalized to the main beam ($I/I_{MB}$). The $z$ value represents the vertical location in the sample. The error bars in (a) and (e) indicate one standard deviation from counting statistics.}
\label{data}
\end{figure}

On top of the standard collimation, a vertical slit is placed in position as shown in Fig.\ \ref{SlicingGeometry}(c). The addition of this single vertical slit in front of the sample focuses our ribbon-shaped neutron beam into a point source, located at the slit itself. The distribution of the neutrons from the source is not completely uniform along the $y$-direction. This vertical variation is due to the fact that the monochromator of AND/R, which is a wavelength selector, is composed of seven separate pyrolytic graphite elements arranged vertically (Dura \emph{et al.}, 2006). The gaps between these elements are visible in Fig.\ \ref{data}(b), a color plot of the intensity from the main neutron beam on the detector with the vertical slit in place. Here only the scattering from three fingers is measurable with the given aperture widths and positions. 

Fig.\ \ref{data}(c) is the raw scattering intensity at \mbox{$x=-0.75$ mm} with a vertical slit of \mbox{$1$ mm} width. In contrast, Fig.\ \ref{data}(d) is the raw scattering data from the same state with the vertical slit wide open (\mbox{$20$ mm}). Comparing these two plots, it is clear that the inhomogeneity of the Bragg peak is revealed by the tight vertical slit.

In order to better visualize the orientational disorder along the length of the flux lines, we decompose the scattering intensity on the areal detector into several sections along the $y$ axis. The number of resolvable sections depends on the vertical collimation. By back-projecting all possible paths of the neutrons from a given $y$-pixel on the detector to the source, we can map the detector regions onto the corresponding vertical locations in the sample ($z$ value). In order to correct for the non-uniform distribution of the main beam, each section is divided by the corresponding main beam intensity ($I/I_{MB}$). Fig\ \ref{data}(e) shows the (corrected) integrated intensity versus vertical position in the sample (\mbox{$z=0$ mm} corresponds to the bottom of the sample) versus the rocking angle $\omega$. In this way we map out both the horizontal and vertical variations of the Bragg peak shape. Fig.\ \ref{data}(a) and Fig.\ \ref{data}(e) demonstrate that the vortex lattice is spatially defective throughout the entire the sample. 

\section{Summary and Conclusions}

The high resolution and ribbon-beam geometry of AND/R have allowed us to probe the spatial dependence of the vortex matter structure transverse to the flux lines in a Nb single crystal. By adding vertical pinhole collimation, we are able to simultaneously measure the orientational disorder along the length of the flux lines in addition to the scattering along the azimuthal and radial directions. The neutron diffractometer provides us with a new approach to map out the complete structure of a vortex lattice and enables us to locate the optimum position inside the sample in searching for the Bragg glass phase. Spatial variations of the orientational order are observed in the lattice structure throughout this Nb single crystal. The technique could easily be adapted for the spatial mapping of other three dimensional structures with suitable geometry.

\section{Acknowledgements}

This research is supported by the U.S. Department of Energy, Office of Basic Energy Sciences, Division of Materials Sciences and Engineering under grant DE-FG$02-07$ER$46458$. NCNR is supported by the U.S. Department of Commerce.  We would like to acknowledge the efforts of William R. Clow and Julia Scherschligt of the NCNR in providing the highly specialized sample environment required to make these measurements. The programs from the reflpak suite were used for elements of the data reduction and analysis (Kienzle \emph{et al.}, 2002 - 2006). X.W. and H.A.H. would like to thank D. Hudek for allowing access to his X-ray diffraction machine.  X.W. and H.A.H. would like to acknowledge support from the Galkin Fund at Brown University.

\newpage

\end{document}